\documentclass{PoS}

\newcommand{\vt}[1]{\mbox{\boldmath{$#1$}}}

\title{Primordial magnetic field constrained from CMB anisotropies,and its generation and evolution before, during and after the BBN}

\ShortTitle{Primordial magnetic field constrained from CMB}

\author{\speaker{Dai G. Yamazaki}\thanks{A footnote may follow.}\\
        Department of Astronomy, Graduate School of Science, University of Tokyo, 7-3-1 Hongo, Bunkyo-ku, Tokyo, 113-0033, Japan\\
        E-mail: \email{yamazaki@th.nao.ac.jp}}

\author{Kiyotomo Ichiki\\
        Research Center for the Early Universe, University of Tokyo, 7-3-1 Hongo, Bunkyo-ku, Tokyo, 113-0033, Japan\\
        E-mail: \email{ichiki@resceu.s.u-tokyo.ac.jp }}

\author{Toshitaka Kajino\\
        Division of Theoretical Astronomy, National Astronomical Observatory Japan, 2-21-1, Osawa, Mitaka, Tokyo, 181-8588, Japan\\
        E-mail: \email{kajino@th.nao.ac.jp}}

\author{Grant J. Mathews\\
        Center for Astrophysics,
Department of Physics, University of Notre Dame, Notre Dame, IN 46556, U.S.A.
\\
        E-mail: \email{gmathews@nd.edu}}

\abstract{The primordial magnetic field (PMF) can strongly affect the cosmic microwave background (CMB) power spectrum and the formation of large scale structure.
In this presentation, we calculate the CMB temperature anisotropies generated by including a power-law magnetic field at the photon last scattering surface (PLSS). 
We then deduce an upper limit on the primordial magnetic field based upon our theoretical analysis of the power excess on small angular scales. 
We have  taken into account several important effects such as the modified matter sound speed in the presence of a magnetic field.
An upper limit to the field strength of $|B_\lambda|\lesssim$ 4.7 nG at the present scale of 1 Mpc is deduced.  This is obtained by comparing the calculated theoretical result including the Sunyaev-Zeldovich (SZ) effect with recent observed data on the small scale CMB anisotropies from the Wilkinson Microwave Anisotropy Probe (WMAP), the Cosmic Background Imager (CBI) and the Arcminute Cosmology Bolometer Array Receiver (ACBAR). 
We discuss several possible mechanisms for the generation and evolution of the PMF before, during and after the BBN.}

\FullConference{International Symposium on Nuclear Astrophysics --- Nuclei in the Cosmos --- IX\\
		 June 25-30 2006\\
		 CERN, Geneva, Switzerland}

\begin{document}
\section{Introduction}
A possible existence of primordial magnetic field (PMF) is an important consideration in modern cosmology.
Recently the origin of the PMF has been studied\cite{bamba04,QCD,Takahashi05,Hanayama05,Siegel:2006px}, and several semi-analytic and numerical studies \cite{mack02,valle04,lewis04,Kosowsky04,yamazaki05a,Dolgov05,yamazaki05b,yamazaki06a,yamazaki06b} also indicate that the effect of the PMF is one of new physical process in earyl Universe which could affect the structure formation on small angular scales for higher $l$. 
However, there is  still no satisfactory understanding of the effects on the PMF, especially on the origin and connection with the observable magnetic field $0.1-1.0\mu$ G in the clusters of galaxies\cite{Clarke:2000bz,Xu:2006rb}. There is a discrepancy of the magnetic field between theory and observation in galactic cluster scale.

Temperature and polarization anisotropies in the CMB
 provide very precise information on the physical processes in operation during the early Universe.
However, the best-fit cosmological model to observed CMB data sets
has recently indicated a potential discrepancy between theory and observation at higher
multipoles $l \ge 1000$.
This discrepancy is difficult to account for by a simple tuning of cosmological parameters.
One possible interpretation of such excess power at high multipoles
is 
the thermal SZ effect \cite{sunyaev80}. 
However, it has not yet been established conclusively that this is the only possible interpretation of the small scale power \cite{Aghanim:2001yu}. 
Indeed, the best value of the matter fluctuation amplitude, $\sigma_8$, to fit
the excess power at high multipoles is near the upper end of the range of 
the values deduced by the other independent methods \cite{Bond:2002tp,Komatsu:2002wc}. Toward the solution of this problem, it has recently been reported that a bump feature in the primordial spectrum way gives a better explanation for both the CMB and matter power spectra at small scales \cite{Mathews:2004vu}.

In order to solve these problems, we study the PMF by likelihood analysis with the Markov Chain Monte Carlo(MCMC) method using data of the Wilkinson Microwave Anisotropy
Probe (WMAP)\cite{Bennett:2003bz}, the Cosmic Background Imager (CBI)\cite{Readhead:2004gy} and the Arcminute Cosmology Bolometer Array Receiver (ACBAR)\cite{Kuo:2002ua} 
We then discuss several possible mechanisms for the generation and evolution of the PMF before, during and after the BBN.
\begin{figure}
\includegraphics[width=1.0\textwidth]{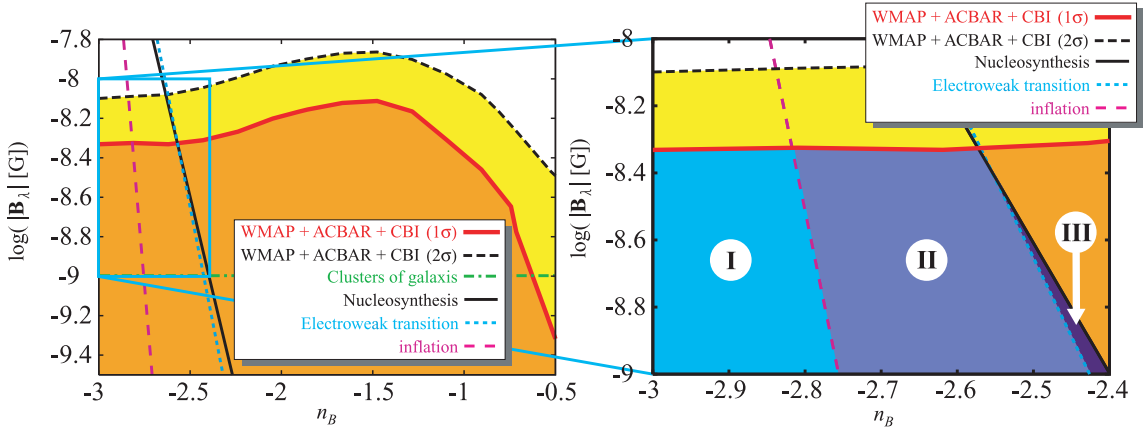}
\caption{Results of the MCMC method constrained by the  WMAP, ACBAR, and CBI data\cite{yamazaki06b}. 
Excluded and allowed regions at the $1\sigma$(68\%) C.L. and $2\sigma$(95.4\%) C.L.~ are shown
 in the two parameter plane of $|B_\lambda|$ vs. $n_B$. 
 Solid and dotted curves are for $\Delta\chi^2=$2.3 and 6.17, respectively. The green dash-dotted horizontal line displays the lower limit of the field strength $B_\lambda=1$ nG deduced for  the galaxy cluster scale at the PLSS \cite{Clarke:2000bz,Xu:2006rb}. The black-solid, skyblue-dotted, and pink-dashed lines are the upper limit of the produced PMF at the big-bang nucleosynthesis, the electroweak transiton, and the inflation epochs, respectively\cite{caprini02}.
 If the PMF is produced at the epochs of big-bang nucleosynthesis, 
  the electroweak transition, or inflation,  respectively, 
  the region of I+II+III, II+III, or III on the right hand side
  are allowed by these constraints on the PMF for the galaxy cluster 
  scale at the PLSS and the MCMC method with WMAP, ACBAR, and CBI data.
 }
\label{fig1}
\end{figure}
\section{Limits on the PMF}
 In Fig.~\ref{fig1} we show results\cite{yamazaki06b} of our MCMC analysis using  the WMAP , ACBAR and CBI data in the two parameter plane $|B_{\lambda}|$ vs. $n_B$, where $|B_{\lambda}|$ is 
 the comoving mean-field amplitude of the PMF obtained
 by smoothing over a Gaussian sphere of comoving radius $\lambda$($\lambda=1$Mpc in this paper), and $n_B$ is the spectral index of the power spectrum of fluctuations from a homogeneous and isotropic distribution of the PMF.
The 1$\sigma$(68\%)
and 2$\sigma$(95.4\%)
 C.L.~ excluded regions are bounded above by the thick curves as shown. 
 We could not find a lower boundary of the allowed region at the 1$\sigma$ or 2$\sigma$  confidence level. 
Note, however,  that we find a very shallow minimum with a reduced $\chi^2 \simeq 1.08$.
From Fig.~\ref{fig1} we obtain an upper limit to the strength of the PMF at the 1$\sigma$(95.4\%) C.L.~of
\begin{eqnarray}
|B_\lambda|\ \lesssim 7.7\ \mathrm{nG}\ \mathrm{at\ 1Mpc}~~.
\end{eqnarray}
This upper limit is particularly robust as we have considered all effects on the CMB anisotropies, i.e. the SZ effect and the effects from both the scalar and vector modes on the magnetic field, in the present estimate of $|B_\lambda|$ and $n_B$.
\section{PMF Generation and Evolution}
In this section we discuss a multiple generation and evolution scenario of the cosmological primordial magnetic field that is motivated by the results of the present study.

To begin with we adopt the following three constraint conditions:

\begin{enumerate}
\item A PMF strength $|\vt{B}_\lambda| \lesssim$7.7nG(1$\sigma$) at 1Mpc as deduced above by applying  the MCMC method to the  WMAP, ACBAR, and CBI data.

\item The magnetic field strength in galaxy clusters is $0.1\mu$G $<\ |B_{CG}|\ <\ 1\mu$ G \cite{Clarke:2000bz,Xu:2006rb}.   Hence, if the isotropic collapse is the  only process which amplifies the  magnetic field strength, the lower limit to the PMF is $\sim 1-10$ nG for the PLSS.

\item The gravity wave constraint on the PMF from Caprini and Durrer (2002)\cite{caprini02}of 0.1 Mpc, however our $\lambda$ is for 1Mpc. Thus, inclinations of lines in Figs. \ref{fig1} are smaller than Caprini and Durer (2004)\cite{caprini02}.
The big-bang nucleosynthesis of the light elements depends on a balance between the particle production rates and the expansion rate of the universe. 
Since the energy density of the gravity waves $\rho_{\mathrm{GW}}$
contributes to the total energy density.  However,  $\rho_{\mathrm{GW}}$ is constrained so that  the expansion rate of the universe does not spoil the agreement between the theoretical and observed light element abundance constraints for 
deuterium, $^3$He, $^4$He, and $^7$Li\cite{Maggiore:2000gv}. 
\end{enumerate}

Figures~\ref{fig1} summarize the various  constraints on the PMF by these three conditions. 
The region bounded by the  upper red-solid curve is constrained by condition (I) as indicated, the green-dash-dotted horizontal line corresponds to  the lower limit to the PMF from condition (II), and the Black-solid, sky-blue-dotted, and pink-dashed lines, respectively, are the upper limit of the produced PMF from big-bang nucleosynthesis, the electroweak transition, and the inflation epoch.
The right hand side of figure. \ref{fig1} shows the allowed or excluded regions according to these multiple constraints depending upon  when the PMF was generated:

(A). Nucleosynthesis: I+II+III region
\begin{eqnarray}
1.0\ \mathrm{nG}\ \lesssim\ |B_\lambda|\ \lesssim\ 4.7\mathrm{nG} \nonumber\\
-3\ \lesssim\ n_B\ \lesssim\ -2.40 \nonumber
\end{eqnarray}

(B). Electroweak transition: II+III region
\begin{eqnarray}
1.0\ \mathrm{nG}\ \lesssim\ |B_\lambda|\ \lesssim\ 4.7\mathrm{nG} \nonumber\\
-3\ \lesssim\ n_B\ \lesssim\ -2.43 \nonumber
\end{eqnarray}

(C). Inflation: III region
\begin{eqnarray}
1.0\ \mathrm{nG}\ \lesssim\ |B_\lambda|\ \lesssim\ 4.7\mathrm{nG} \nonumber\\
-3\ \lesssim\ n_B\ \lesssim\ -2.76 \nonumber
\end{eqnarray}

Obviously,  the upper limits on both $|B_\lambda|$ and $n_B$ become more stringent if the PMF is produced during an earlier epoch.  
Moreover, the  limits we deduce are the strongest constraints on the PMF that have  yet been determines. However, we caution that the evolution of the generation of the PMF during the LSS epoch is not well understood. Thus, if there are other effective physical processes for the generation and evolution of the PMF during the formation of LSS, our lower limit of the PMF parameters may decrease.
 In order to constrain the PMF parameters accurately, we should study the PMF not before but after the PLSS.
\section{Conclusion}
In order to constrain the PMF, we evaluate likelihood function of the WMAP TT, CBI, and ACBAR data in a wide range of parameter of the magnetic field strength $|B|_\lambda$ at 1 Mpc and the power-law spectral index $n_B$,  along with six cosmological parameters, the Hubble parameter, the baryon and cold dark matter densities, the spectral index and the amplitude of primordial scalar fluctuation,
and the optical depth, in flat Universe models, using the technique of the MCMC method.

For the first time we have studied  scalar mode effects of the PMF on the CMB\cite{yamazaki06b}.
We have confirmed numerically without approximation that the excess power in the  CMB at higher $l$ can be explained by the existence of a PMF.
For the first time a likelihood analysis utilizing the  WMAP, ACBAR and CBI data with a MCMC method has been applied to constrain the upper limit on the strength of the PMF to be 
\begin{eqnarray}
|B_\lambda| < 7.7nG ~~.\nonumber
\end{eqnarray}
We have also considered three conditions on the generation and evolution of the cosmological PMF:
1) our result;
2) the lower limit of the PMF from the magnetic field of galaxy clusters;
 and 3) the constraint on the PMF from gravity waves. 
 COmbining these, we find the following concordance region for the  the PMF parameters;
\begin{eqnarray}
1~\mathrm{nG}< |B_\lambda| < 4.7~ \mathrm{nG}~~,\ \ -3.0< n_B < -2.4~~.\nonumber
\end{eqnarray}

The PMF also affects the formation of  large scale structure.  For example, magnetic pressure delays the gravitational collapse. It is thus very important to constrain the PMF as precisely as possible. 
If we combine our study and future plans to observe the CMB anisotropies and polarizations for higher multipoles $l$, e.g.~via the {\it Planck Surveyor}, we will be able to constrain the PMF more accurately, and explain the evolution and generation of the magnetic field on galaxy cluster scales along with  the formation of the LSS.

%\begin{figure}
%\includegraphics[width=.6\textwidth]{f2.eps}
%\caption{Excluded and allowed regions at the $1\sigma$(68\%) C.L.
% and $2\sigma$(95.4\%) C.L. 
% in the two parameter plane $|B_\lambda|$ vs. 
% $n_B$ obtained by the MCMC method with the WMAP, ACBAR, and CBI data. 
% $|B_\lambda|$ is the primordial magnetic field strength 
% and $n_B$ is the power-law spectral index. 
% The solid and dotted curves are for $\Delta\chi^2=$2.3 and 6.17, respectively.
%  The  black-solid, skyblue-dotted, and pink-dashed lines are 
%  the upper limit of the produced PMF at the big-bang nucleosynthesis, 
%  the electroweak transition, and the inflation epochs, respectively 
%  (Caprini and Durrer 2002). 
%  If the PMF is produced at the epochs of big-bang nucleosynthesis, 
%  the electroweak transition, or inflation,  respectively, 
%  the region of I+II+III, II+III, or III 
%  are allowed by these constraints on the PMF for the galaxy cluster 
%  scale at the PLSS and the MCMC method with WMAP, ACBAR, and CBI data.}
%\label{fig2}
%\end{figure}

\end{document}